*The Making of a Creative Worldview*
Liane Gabora

Chapter for *Secrets of Creativity*
To be published by Oxford University Press
Edited by Suzanne Nalbantian and Paul Matthews

Address for correspondence:
Liane Gabora
Department of Psychology, Fipke Centre for Innovative Research
3247 University Way
Kelowna BC, Canada V1V 1V7
Email: liane.gabora [at] ubc.ca
Tel: 250-807-9849

Research at the interface between cognitive psychology, neuroscience, and the science of complex, dynamical systems, is piecing together an understanding of the creative process, including how it works, how it can be fostered, and the developmental antecedents and personality traits of particularly creative people. This chapter examines the workings of creative minds, those with the potential to significantly impact the evolution of human culture.

Think of the least creative people you know. Their mental models of reality consist of knowledge, social rules, and norms they've picked up from others through social learning processes such as imitation. To understand this, I focus on the global structure of their minds, or what I call their 'worldviews'. A *worldview* is a mind as experienced subjectively, from the inside; it is a way of *seeing* the world and *being in* the world that emerges as a result of the structure of ones' web of understandings, beliefs, and attitudes. A worldview reveals itself through behavioural regularities in how it expresses itself and responds to different situations.[1]

Research on the personality of creative individuals suggests that they would be likely to exhibit high energy, curiosity, openness, confidence, rebelliousness, and a propensity to deeply immerse themselves in projects they feel passionate about to the exclusion of all else.[2] Uncreative people take in information through social learning (learning from other people), which psychologists sometimes contrast with *individual learning*, which involves learning that occurs through direct experience in the world, outside a social context. But creative people do something that goes beyond both social and individual learning. Their worldviews are not merely compilations of the knowledge, social rules, and norms that they have picked up from the world around them; they have reflected on and rehashed this information to put their own spin on it, through processes that Piaget referred to as assimilation and accommodation.[3] *Assimilation* involves fitting new information into one's existing web of understandings, whereas *accommodation* is the complementary process of restructuring one's existing web of understandings to make sense of the new information.[4] The global structure of their worldview is not like a collection of discrete and separate pebbles; it is interwoven, malleable, and ever-changing, more like a kaleidoscope or fractal screensaver.

Simply put, while the structure of the worldview of an uncreative person reflects *what they've been told*, the structure of the worldview of a creative person reflects *what they've done with what they've been told:* their flights of fantasy and imagination, as well as symbol manipulation and deductive and abductive processes. For simplicity, we will call the worldview of an uncreative person a *socially-made worldview* and the worldview of a creative person a *self-made worldview*. Of course, in reality, almost everyone falls somewhere in between these two extremes.

Like a biological organism, a worldview is *self-organizing* and *self-mending*, and this is particularly true of a self-made worldview.[5] Such brains explore and play with knowledge and experiences code into memory, looking at them from different perspectives, such that over time they tend to form a coherent web and take a form that may bear little resemblance to the form in which they were originally acquired. The greater the tendency of a worldview to self-organize, the more likely it is to arrive at a structure that deviates from the status quo, and this internal upheaval can be demanding, not just cognitively but also emotionally. The self-made worldview individual may be more likely to been seen as unusual,[2] or even diagnosed with an affective disorder[6], but has greater potential to produce something—whether it be an idea or artifact—that the world will deem new and original, i.e., creative.

## Psychological Entropy and The Gap

Someone with a self-made worldview is particularly inclined to experience what Torrance[7] calls a *gap*: a question or problem, or sense of incompletion, or feeling of curiosity, or emotion in need of expression, which may arise spontaneously, or slowly over the course of years, and be either trivial or of worldly consequence. The gap that precipitates creative thinking has been described as a relatively chaotic cognitive state[8] that is generally accompanied by a lingering feeling that compels the exploration and expression of ideas.**Error! Bookmark not defined.** The gap functions as an attractor state, luring the creative individual in to explore an area of disconnect.

The notion of the gap can be framed in terms of the concept of *entropy* a term which comes from thermodynamics and information theory and refers to the uncertainty or disorder in a system. Self-organizing systems, such as worldviews, continually interact with and adapt to their environments to minimize internal entropy. Open systems, e.g., living organisms, must capture energy (or information) from their environment to maintain semi-stable, far-from-equilibrium states of low entropy. Such self-organizing systems, modify their contents and adapt to their environments to minimize entropy. Hirsh and colleagues[9] use the term *psychological entropy* to refer to psychological uncertainty, which they claim humans attempt to keep at a manageable level. Arenas of life where there is conflict, confusion, or uncertainty attract our attention; we automatically zero in on them, because they are associated with changes in this psychological entropy. Creativity may begin with detection of high psychological entropy, which provokes uncertainty and arousal. [1] The detection of the psychological entropy gap in turn signals to the individual that a particular arena of understanding could benefit from self-organized restructuring.

**Chaining: The Capacity for One Thought to Trigger Another**

How does a mind self-organize and restructure itself? While other species can come up with creative ways of doing things, we humans are uniquely able to adapt ideas to our own needs, tastes, and situations, and in so doing, our understanding of our world, and our own place in it, is constantly shifting and unfolding.

My colleagues and I have argued that the extraordinary creativity of the human mind is due to the onset of two uniquely human cognitive abilities.[10] The first, which was marked by the appearance of the earliest stone tools over two million years ago, was the onset of the capacity for one thought to trigger another thought.[11] Donald called this the *self-triggered recall and rehearsal loop* because it enabled our early ancestors to rehearse and refine skills, recall and reflect upon the past, and imagine the future. One form of self-triggered recall is the *chaining* of thoughts and actions into streams of free-association, critical reflection, or complex behavioral sequences.[12] Self-triggered recall enabled our early hominid ancestors to recursively consider high psychological entropy material from new contexts until it was sufficiently restructured that a new idea or perspective was achieved, and arousal dissipated.

It can be argued that self-triggered recall became possible because a natural consequence of having a more *fine-grained* memory is larger *working memory*, and moreover, that larger working memory in and of itself is not useful unless it goes hand-in-hand with more fine-grained memory.[10] The benefit of being able to hold multiple items or properties in mind at the same time comes from being able to use them in sequenced actions or refine how they relate to one another in streams of thought. Thus, our ancestors became capable of self-triggered recall by connecting the dots between items in memory previously assumed to be unrelated. They went from being stimulus-response machines with simple responses, to being able to respond more flexibly, and reflect on their responses. The hypothesis that cultural evolution was made possible by the onset of the capacity for self-triggered recall processes such as chaining was tested in EVOC (for EVOlution of Culture), an agent-based computer model of cultural evolution.[13] In EVOC, it was demonstrated that chaining increases the mean fitness (i.e., usefulness) and diversity of cultural outputs. Findings in EVOC support the hypothesis that the ability to chain ideas was what transformed a culturally static society into one characterized by open-ended novelty.[14] The greater the extent to which one 'remakes' one's worldview by using chaining to weave new understandings of the past or speculate about the future, the less one remains alert to the 'here and now', but if something *does* manage to attract attention, it is more likely to be thoroughly processed before settling into a particular interpretation of it.

**Contextual Focus: The Capacity to Shift between Modes of Thought**

One of the possible models for the evolution of mind involves a second human cognitive ability that is posited to have evolved approximately 60,000 years ago, giving rise to what Mithen refers to as the 'big bang' of human creativity in the Middle-Upper Paleolithic. [15] It is referred to as *contextual focus* because it entails the capacity to adapt ones' mode of thought to the context by focusing or defocusing attention. [16] Contextual focus enabled our ancestors to shift spontaneously between *convergent* and *divergent* modes of thought. Convergent or analytic thought is conducive to stifling associations, and reserving mental effort for symbol manipulation and detecting relationships of *causation* when engaged in mentally demanding analytic task. *Divergent* or *associative* thought is conducive to forging new associations when "stuck in a rut", conducive to detecting relationships of *correlation*, and simply letting ones' mind wander. In associative thought, the interconnections between concepts and ideas become more fluid and malleable. [17]

Whereas chaining enabled the connecting of *closely* related items in memory, contextual focus enabled the forging of *distant* connections (e.g., drawing analogies), and sophisticated creative expression (e.g., chiseling of a representation of an animal into rock). Divergent thought enabled distant connections to be glimpsed, and convergent thought enabled them to be polished into their final form. Thus, it seems reasonable that the ability to utilize both associative and analytic thought, and to know *when* to use them, is important to the formation of a self-made worldview.

It has been proposed that while convergent thought involves using concepts in their most stereotypical, compact, undifferentiated form, i.e., *sticking to their most conventional contexts*, divergent thought is characterized by exploring the broader 'halo' of potentiality surrounding concepts, *i.e., the new meanings or feelings that arise when concepts are conceived of in particular, often unconventional contexts.* [18] Divergent thought is not a matter of generically *loosely* expanding ones' sphere of associations, but of making broad use of contextual information to make associations in *constrained* (though potentially obscure) ways.

During creative thought, there are neural mechanisms (outlined below) that enable the shifting between analytic and associative modes of thought occurring on the level of groups of collectively co-spiking neurons or *neural cliques*. There will be some neural cliques that *would not* be included in a cell assembly if one were in an analytic mode of thought, but *would* be included if one were in an associative mode of thought. We have referred to these neural cliques as *neurds*. [19] Neurds respond to features that are peripherally relevant to the current thought. Neurds do not reside in any particular region of memory; the subset of neural cliques that count as neurds shifts constantly, and is defined by context. For each different situation, there would be a different group of neurds.

The explanation of creativity proposed here follows from the the concept of contextual focus and from the well-established phenomenon that activation of an abstract or general concept causes activation of its instances through a mechanism such as spreading activation. [20] Those neural cliques that respond to more general or abstract aspects of a situation offer a straightforward mechanism for contextual focus. In associative thought, with more aspects of a situation taken into account, more neural cliques are recruited, including those that respond to highly specific features, those that respond to more abstract features, and those *they* activate through spreading activation. Activation flows from the specific instance to the abstract concepts it instantiates, and then to other instances of those abstract concepts. Thus, the neurds concept provides a means of referring to neural cliques that respond to features of these other instances that were not present in the original instance.

It is likely that most of the time, the activated cell assemblies do not include neurds. However, when there is a problem to be solved, broad activation causes more neural cliques to be recruited, including neurds, and thought becomes more associative. This enables each successive thought to stray far from the previous one, yet retain a thread of continuity. The worldview is traversed quickly and penetrated deeply, such that new creative connections are more likely to be made.

As with chaining, contextual focus contributes to the making of a self-made worldview through the forging of unusual and often distinctively personal associations. Conversely, the self-made worldview individual experiences more psychological entropy, and thus there is a greater perceived need for restructuring, and thus for contextual focus. When psychological entropy signals a gap in a worldview, divergent thought ensues in an attempt to forge new connections and coming up with ideas that mend their fractured worldview. However, since the new ideas conceived in a divergent mode of thought are not necessarily immediately implementable, or even intelligible, they must be reflected upon or honed from different perspectives in a more sober, convergent mode of thought. In short, the self-made worldview subjects the fruits of one mode of thought to another mode of thought thereby achieving a more nuanced web of associative structure, and this is reflected in creative behavior and artifacts.

Like chaining, contextual focus has also been modeled using the EVOC computer model.[14] The mean fitness of actions across the society increased with contextual focus, and contextual focus was also particularly effective when the fitness function changed. These findings support the hypothesized usefulness of contextual focus in problem solving, seeing something from a new perspective, adapting to new or changing environments, and sparking insight. The process of putting chaining and contextual focus to work in the generation of a creative output is sometimes referred to as *honing*.[1]

## Honing and the Imagination

Honing involves viewing something in a new context, which leads to a more nuanced view of it. For example, if you consider your new dog from the context, 'what would my mother think,' your conception of your dog now encompasses your mother's views on barking and shedding. (For a precise, mathematical description of this process, see Aerts et al.)[21] This more nuanced view of it suggests a new context to consider it from, and so forth, until psychological entropy reaches an adequately low level, the gap is filled, and the worldview is less fragmented. Chaining is sufficient for little-c 'everyday' creative ideas, while contextual focus is more likely to come into play in the generation of big-C, history-making creative contributions. Through immersion in a creative task, a more stable image of both the world and one's own relation to it comes into focus.[22] Honing may make use of *imagination*, the ability to create internal images or ideas by mentally combining previous experience and knowledge. Since a self-made worldview is in a process of constant renewal, it is likely to experience cognitive states that feel ill-defined, and in need of honing. Indeed, a newly hatched idea may not be completely comprehensible even to the creator.

## Insight

The process of honing what one is thinking about by reiteratively viewing it in a new context and thereby getting a new take on it can lead to insight. *Insight* can be defined as a sudden new representation of a task, a realization of how to go about it,[23] or conscious cognitive reorganization concluding in a new and useful but non-obvious idea, interpretation, or understanding.[24] Although insight is often experienced as a sudden and emotionally uplifting 'aha moment', studies have shown it to be the culmination of a series of brain states and processes operating at different time scales.[25]

While some attribute the unpredictability of insight to the blind, trial-and-error nature of idea generation, another explanation is that it marks a cognitive phase transition arising due to *self-organized criticality*,[26] a phenomenon first studied by Bak, Tang, and Weisenfeld.[27] Bak and his colleagues used the term 'criticality' because, through simple local interactions, complex systems tend to find a critical state, poised at the cusp of a transition between order and chaos, from which a single small agitation occasionally exerts a disproportionately large effect. The signature of self-organized criticality is an inverse power law relationship between the degree of a critical event and the frequency of critical events of equal degree or importance. Just as most of our thoughts are inconsequential, once in a while one thought that triggers another, a chain reaction can causie an 'avalanche' of reconceptualizations and culminate in a dramatically altered understanding of something.

Like other systems that exhibit self-organized criticality, a creative mind may function within a regime midway between order (systematic progression of thoughts) and chaos (everything reminds one of everything else).[17] Such a mind may not simple be *poised* in this regime, it may use contextual focus to *stay* in this regime: systematic analysis when life gets complicated, and letting the mind wander when things get dull.

The proposal that insight arises through self-organized criticality is consistent with findings that large-scale creative conceptual change often follows a series of small conceptual changes,[28] and with evidence that power laws and catastrophe models are applicable to innovation.[29] It is also consistent with findings that, although often preceded by intensive effort, tends to come suddenly and with ease.[30] Self-organized criticality gives rise to structure that exhibits sparse connectivity, short average path lengths, and strong local clustering. Other indications of self-organized criticality include long-range correlations in space and time, and rapid reconfiguration in response to external inputs. There is evidence of self-organized criticality in the human brain, e.g., with respect to phase synchronization of large-scale functional networks.[31] There is also evidence of self-organized criticality at the cognitive level; word association studies have shown that the organization of concepts in the mind exhibits signature properties of self-organized criticality such as sparse connectivity and strong local clustering.[32] Thus, semantic networks exhibit the sparse connectivity, short average path lengths, and strong local clustering characteristic of self-organized complexity and 'small world' structure.[33]

## Associative Memory: The Cradle of Creative Ideas

To see how the notion of a self-made worldview can be grounded in the cognitive architecture, we begin by examining some key attributes of associative memory. The structure of memory is critically important for creativity.**Error! Bookmark not defined.**

First, as pointed out some time ago by Kanerva the architecture of memory is *sparse*, in the sense that although a human brain can have approximately 100 billion neurons, the number of possible items that can be encoded is far greater.[34]

Second, memories exhibit *coarse coding*.[35] Human memories are encoded in neurons that are sensitive to ranges (or values) of features. Coarse coding is an organization that results in a neuron responding maximally to a particular microfeature, and responding to a lesser extent to similar features. For example, neuron *A* may respond preferentially to a certain sound, while its neighbor *B* responds preferentially to a slightly different sound, and so forth. However, although *A* responds maximally to one sound, it still responds, to a lesser degree, to another similar sound.

When an item in memory is *distributed* across a cell assembly of neurons that are sensitive to particular high-level or low-level properties, this ensures the feasibility of forging associations amongst items that are related, perhaps in a surprising but useful or appealing way.[36] A given experience activates not just one neuron, nor every neuron to an equal degree, but spreads across members of an assembly. The semantic content of a memory arises due to its overall pattern of activation, as opposed to any single neuron. The same neurons get used and re-used in different—and sometimes novel—capacities.[37] As such, the way items are encoded in memory takes into account how they are related, even if this relationship has never been consciously noticed.

Unlike a conventional computer, in which information is accessed according to its address (regardless of its content), human memory is *content addressable*;[1] there is a systematic relationship between stimulus content and the cell assemblies that encode it. This ensures that the brain naturally brings to mind items that are related in obvious or unexpected ways to what is being experienced, i.e., memory items are awakened by stimuli that are similar or 'resonant'.[38] When you encounter something in the external world, say something fuzzy, neurons that respond to attributes of that thing are activated; e.g., neurons that encode memories of previously encountered fuzzy things are reactivated. Although the correlation between them may have never been explicitly noticed, the fact that their distributions overlap means that one can evoke the other.

Thus, memory is sparse, distributed, course coded, content addressable. Neurons operate in parallel and are often reciprocally connected, such that large sets of neurons mutually constrain one another in a way that is sensitive to both structure and content. This enables them to integrate information from many different sources, exhibit context sensitivity, and respond simultaneously, in graded fashion, to multiple constraints.[39] These features, taken together, mean that neural networks are able to abstract a prototype, fill in missing features of a noisy or incomplete pattern, or create a new pattern on the fly that is more appropriate to the situation than anything it has ever been fed as input.[40]

This cognitive structure, in conjunction with a self-made worldview that organizes itself in response to psychological entropy, *is* the key to our creativity. If memory were not coarse coded and distributed, then there would be no overlap between items that share features, and thus no means of forging associations between them. If memory were not content addressable, then these associations would not be relevant and meaningful.[41] Content addressability makes it possible to hit upon the right thing at the right time such that memory need not be searched or randomly sampled for relevant creative associations to be made. If memory were not sparse—if all neurons were crowded into some small region of feature space (such as, say, detection of different texture)—we would not be able to make distinctions across the vast range of sensory qualities that we can.

---

[1] Note that content addressability is increasingly common in computers.

Memory does not work through a simple, verbatim retrieval process; items are never re-experienced in exactly the same form as when first encoded. Recalled items are unavoidably colored, whether subtly or strongly, by what has been experienced since encoding, and can be spontaneously re-assembled in a way that relates to the task at hand.[42] The contents of memory are in constant flux, undergoing construction and reconstruction, through modification of the strengths and patterns of connection amongst neurons.[40] This is how a worldview gets restructured.

Memories become *integrated* through the recruitment and *binding* of overlapping populations of these distributed, content-addressable neurons, a process that involves the hippocampal–medial prefrontal circuit.[43] Events that take place close in time are more likely to be bound in this way.[44] This binding is the key to the forging of new ideas from existing ones. Retrieving a specific, singular memory is essentially impossible, as a memory is not stored in seclusion, but is distributed and also experientially shaped. Because information is encoded in a distributed manner across neurons assemblies, the meaning of an item in memory is in part derived from the meanings of other representations that excite similar constellations of neurons (i.e., constellations that are associated).

## Chaining at the Level of Neural Cell Assemblies

We have seen that chaining was made possible through the onset of richer encodings of items in memory as a result of the larger, more interconnected brain believed to have been made possible due to enlarged cranial capacity. At the level of neural cell assemblies, this meant that more microfeature specific neurons were available to participate in the encoding of memories, thus more features of any particular experience were encoded. Therefore, more distinctions could be made, there were more retrieval routes by which one item could interact with another, and more possibilities for forging novel relationships between current and past experiences. In this way, the proclivity for a self-made worldview may begin with the tendency to encode items in memory in greater detail (i.e., the more properties of the stimulus are recorded) than more routes by which one thought can evoke another, and in turn, a more nuanced worldview.

## Contextual Focus and Insight at the Level of Neural Cell Assemblies

It has been proposed that the shifting between associative and analytical thought in contextual focus is carried out through recruitment and decruitment of neural assemblies called *neurds*.[18] Neurds are the cell assemblies that are *not* activated in analytic thought, but that *are* in associative thought (see also Ellamil, Dobson, Beeman, & Christoff,[45] and Yoruk & Runco)[46]. Neurds respond to properties that are of minimal relevance to the current thought, and are thus more distantly associated. Neurds do not reside in any particular region of memory; the subset of cell assemblies that count as neurds shifts depending on the situation. Every different perspective one takes toward a particular thought, idea, or concept recruits a different set of neurds.

In associative thought, diffuse activation causes more cell assemblies to be recruited, including neurds, enabling one thought to stray far from the preceding one while still retaining a thread of continuity and overlap. Overlapping mental representations have greater potential to be become bound together, resulting in the uniting of ideas or concepts previously assumed to be unrelated. Insight is the '*aha!*' moment that occurs when a concept and its novel associates are sufficiently activated to cross the threshold of conscious awareness.[47] At the neural level, the merging of thoughts culminating in insight may involve recurrent connections in the hippocampus, particularly when the insight involves generalization and inference triggered by a particular recent experience.[48] Research on the neuroscience of insight suggests that alpha band activity in the right occipital cortex causes neural inhibition of sensory inputs, which enhances the relative influence of internally derived ("non-sensory") inputs, and thus the forging of new connections.[24]

Following insight, the shift to an analytic mode of thought—in order to hone an idea to completion—could be accomplished through decommissioning of the relevant neurds (i.e., silencing of the cell assemblies that are active during associative thought). Findings that the right hemisphere tends to engage in coarser semantic coding and have wider neuronal input fields than the left[24] suggest that the right hemisphere may dominate during associative thought while the left takes the lead during analytic thought.

Clearly, the greater the extent to which one can tailor the mode of thought to the situation through contextual focus, the more likely one is to have a self-made worldview. In associative thought, the memory is navigated not just deeply, but quickly, and this affects not just the content of current thought, but results in a more integrated worldview, which its proclivities for future thoughts. However, since too much integration is counterproductive (you don't want everything to remind you of everything all the time) the ability to reign in associative thinking is as important as the ability to engage in it in the first place. This theory posits that the greater the ability to shift between modes of thought, the greater the extent to which the structure of one's worldview incorporates, not just contents acquired through social and individual learning, but also their interrelationships, which are the source of creativity. This includes relationships amongst concepts at different levels of abstraction, and considered from differend contexts (e.g., in the context 'color', the poet might draw upon the redness of both fire engines and hearts, despite that in other contexts they would not appear to be related). Creative individuals constantly reexamine and reorganize what they experience, using psychological entropy to guide self-supervised learning and achieve a more nuanced self-made worldview , with creative outputs.

## The Relationship of Insightful Ideas to the Structure of Memory

In some ingeniously designed experiments, Bowers, Farvolden and Mermigis showed that, even when insight appears to come suddenly or 'out of the blue', it is preceded by unconscious processing that is leading the creator toward the response.[49] Having examined how new ideas emerge in associative memory, we have a better understanding of what is taking place during this time at the neural level. The fact that associations come to mind spontaneously as a result of representational overlap and sharing of features means there is no need for memory to be searched or randomly sampled in order to make creative associations. The more detail with which stimuli and experiences are encoded in memory, the greater the degree to which their distributed representations overlap, and the more potential routes by which they can act as contexts for one another and combine. They may have been encoded at different times, under different circumstances, and the relationship between them never explicitly noticed, but some situation could come along and make their relationship apparent. Thus, what gets pulled out of your brain during a creative episode bears some relationship to knowledge and experiences encoded in memory before the creative act took place.

In short, due to the organization of memory, insight is viewed as not primarily a matter of chance or expertise, but as spotting previously unnoticed areas of overlap amongst memories and concepts, becoming newly aware of their shared features.

## Creative Worldviews as the Basic Units of Cultural Evolution

In seeking to unlock the mysteries of creativity, we must also examine what role it plays in the grand scheme of things. Humans participate not only in biological evolution, but also a second evolutionary process—the evolution of culture—in which creativity is the driving force.[50] And the evolution of culture is fueled by creativity, i.e., the creative process enables the unlimited proliferation of cultural outputs.

In attempting to understand  the process by which culture evolves, it is interesting to note that very early life itself went through a stage where it exhibited cumulative, adaptive, open-ended change—it evolved—but through a process that, at least initially, worked quite differently from natural selection. What made this early *biological* evolution possible was the emergence of simple cell-like structures called *protocells* that were self-organizing, self-mending, communally interacting, and self-reproducing.[51] It has been shown that these protocells evolved through process that has been referred to as *communal exchange* because it involved interactions not just within but amongst these self-organizing structures.[52] The process was more haphazard than natural selection but sufficient for cumulative, adaptive, open-ended change.

It has been proposed that the cultural equivalent of a biological organism is a worldview, i.e., what made *cultural* evolution possible is the emergence of a kind of *mind* that is self-organizing, self-mending, communally interacting, and self-reproducing.[53] We have already seen how worldviews are self-organizing and self-refining. They are also communally interacting. Just like the protocells that constituted the earliest forms of life, the internal state of the worldview changes dynamically due to a communal exchange of information with the external world. In turn, an individual shares experiences, ideas, and attitudes with others, thereby influencing the process by which other worldviews will emerge and change.

To the extent that someone interacts—directly or indirectly—with others, the configuration of this individual's worldview influences and is imperfectly reconstituted in others, and is thus passed on. For example, children expose fragments of what was originally the adult's worldview according to the child's own different experiences and bodily constraints, thereby forging unique internal models of the relationship between self and world. It is in this sense that worldviews are not just communally interacting, they are self-regenerating as well. The more information a worldview encodes, as a result of social and environmental interactions and experiences, the more ways it can be configured, and the more likely it is to unfold in a unique way and become a self-made worldview.

Thus, a worldview evolves through communal exchange by interweaving both (1) internal interactions amongst its parts, and (2) external interactions with others. Because unlike biological evolution there is no universal 'self-assembly code' ensuring reliable of replication, a particular belief or idea may play a different role within different worldviews and be colored by the experiences and structures unique to that particular worldview.

One of the more striking implications of communal exchange theory is that it is not the creative products (e.g., artifacts such as tools or works of art, or actions such as gestures or dance steps) that are evolving but the worldviews that give rise to them. The state of a worldview reveals itself in social interaction through behavioural regularities in how individuals both creatively express their worldviews and respond to affordances and perturbations.[1] In turn, these products shape the worldviews of those who appreciate, use, or interact with them. The products themselves are, in a sense, the effluent of worldview evolution.

## Conclusions

Just as the cumulative creativity of biological evolution completely transformed our planet, the cumulative creativity of cultural evolution is completely transforming human society. Although the creative processes underlying cultural evolution may be more strategic n than biological evolution, we still cannot predict where its headed. While we have made considerable headway in understanding how the creative process works, we are no better able to predict the next new gadget that will captivate us or the construction of the novel that will move us to tears. We *can*, however, say something about the kinds of people that will generate these creative outputs. Their worldviews are not compilations of knowledge and social rules; they are self-made worldview s that weave what they are told into a cognitive structure of their own making. Their worldviews will reflect not just their unique proclivities and experiences, but how they reflect upon them and hone them into the ideas that shape our world.

## Acknowledgements

The author thanks the Natural Sciences and Engineering Research Council of Canada for funding (grant 62R06523) and thanks Alexandra Maland for assistance with the manuscript.

## Notes